\documentclass[prl,twocolumn,superscriptaddress,amsmath,amssymb,showpacs]{revtex4-1}

\usepackage{graphicx}
\usepackage{color}
\usepackage[dvips]{epsfig}
\begin{document}

\title{Nonlocal biphoton generation in Werner state from a single semiconductor quantum dot}
\author{H.\ Kumano}
    \email[]{kumano@es.hokudai.ac.jp}
    \affiliation{Research Institute for Electronic Science, Hokkaido University, Sapporo 001-0021, Japan}\author{H.\ Nakajima}
    \affiliation{Research Institute for Electronic Science, Hokkaido University, Sapporo 001-0021, Japan}
\author{T.\ Kuroda}
    \affiliation{National Institute for Materials Science, 1 Namiki, Tsukuba 305-0044, Japan}
\author{T.\ Mano}
    \affiliation{National Institute for Materials Science, 1 Namiki, Tsukuba 305-0044, Japan}
\author{K.\ Sakoda}
    \affiliation{National Institute for Materials Science, 1 Namiki, Tsukuba 305-0044, Japan}
\author{I.\ Suemune}
    \affiliation{Research Institute for Electronic Science, Hokkaido University, Sapporo 001-0021, Japan}

\date{\today}

\begin{abstract}
We demonstrate Werner-like polarization-entangled state generation disapproving local hidden variable theory from a single semiconductor quantum dot. By exploiting tomographic analysis with temporal gating, we find biphoton states are mapped on the Werner state, which is crucial for quantum information applications due to its versatile ramifications such as usefulness to teleportation. Observed time evolution of the biphoton state brings us systematic understanding on a relationship between tomographically reconstructed biphoton state and a set of parameters characterizing exciton state including fine-structure splitting and cross-dephasing time.
\end{abstract}


\maketitle
Quantum biparticle state is the simplest physical system which could exhibit profound quantum-mechanical phenomena such as entanglement between causally independent particles~\cite{EPR1935} and nonlocality ~\cite{Bell64, CHSH69}. These properties are the heart of quantum information and communication technology providing unconditional security~\cite{Horodecki09, Gisin_PRL10}.
For two-qubit pure states, entanglement and nonlocality are equivalent~\cite{Gisin_PL91}.
In practice, however, all the systems are inevitably driven into mixed states because any system is more or less open to its environment and subject to loss and decoherence.
Although the relationship between entanglement, nonlocality, and teleportation fidelity is not fully understood for general two-qubit mixed states~\cite{Buscemi_PRL12, Cavalcanti13}, one of the mixed entangled state, so-called Werner state~\cite{Werner89, add3} has been widely investigated due to its widespread ramifications.
For example, there exists bipartite mixed states which are entangled but do not violate any Bell-type inequalities~\cite{Werner89}, and all the entangled Werner states are useful for teleportation~\cite{Mista_PRA02}.
Moreover, the Werner states can be regarded as maximally entangled mixed states of two-qubit systems whose degree of entanglement cannot be increased by any unitary operations~\cite{Ishizaka_PRA00}.
Therefore generating the Werner states is of significant importance
for practical biphoton sources employed in the field of quantum information and communication technology.

For quantum photon sources with parametric down conversion~\cite{Kwiat_PRL95, Kwiat_PRA99},
signal and idler photons have intrinsic quantum-mechanical correlation.
Biphoton states have been extensively studied with quantum state tomography~\cite{White_PRA01, Rangarajan09, Kumano_OptExp11},
and the Werner state formation was proved by introducing polarization diffusers~\cite{Puentes_PRA07,Zhang_PRA02}.
On the other hand, quantum-dot (QD) photon sources~\cite{Michler09},
despite of their potential feasibility for quasi-deterministic operation,
biphoton generation with quantum correlation is not straightforward.
Lowering symmetry of exciton confinement potential results in anisotropic $e$-$h$ exchange
and brings fine-structure splitting (FSS) in bright exciton states~\cite{Bayer_PRB02}.
Resultant which-path information hinders quantum correlation by breaking superposition
between two decaying paths $H_{XX}H_{X}$ and $V_{XX}V_{X}$
for neutral biexciton (${\rm XX}^0$)-exciton (${\rm X}^0$) cascading process~\cite{Kumano_JNO06}.
So far, in order to suppress the which-path information for a selected QD, electric~\cite{Vogel_APL07, Bennett_NPhys10, Ghali_NComm12}, magnetic~\cite{Stevenson_PRB06}, and strain~\cite{Seidl_APL06, Trotta_PRL12} fields, and spectral~\cite{Akopian_PRL06} or temporal~\cite{Stevenson_PRL08, Young_PRL09} filtering were applied, thence polarization-entangled~\cite{Vogel_APL07, Bennett_NPhys10, Ghali_NComm12,Stevenson_PRB06, Seidl_APL06, Trotta_PRL12, Akopian_PRL06, Stevenson_PRL08} or nonlocal~\cite{Young_PRL09} photon-pair generation has been achieved.
However, biphoton states generated from the QD-based sources are argued basically
from a viewpoint of the state being entangled (or nonlocal) or not,
%
%
and further details on the biphoton states against all the physically possible
\textit{biphoton mixed states} remain elusive yet.

In this letter, biphoton states via the ${\rm XX}^0$-${\rm X}^0$ cascading emission
from a QD are systematically examined based on an analytical density matrix
for the excitonic system given by Hudson et al.~\cite{Hudson_PRL07}.
Density matrix which possesses the full information on the biphoton state
is tomographically reconstructed and directly compared with analytically evaluated one.
Highly symmetric QDs are prepared and give nonlocal biphoton mixed state without any fields
or filtering process~\cite{Kuroda_PRB13}, which enables us to access the time evolution
of the biphoton state in wider time range
between ${\rm XX}^0$ and ${\rm X}^0$ emission.
As a result, we have successfully established the quantum mechanical description
for the biphoton state from a quantum dot emitter as Werner state,
which distinguishes the obtained biphoton state from generally allowed biphoton mixed states.
Fundamental parameters to determine underlying dynamics
in the exciton state in the QD are clarified
and a direction towards the ideal biphoton source is presented.

As a biphoton emitter, we employ unstrained GaAs QD formed on lattice-matched
Al$_{0.3}$Ga$_{0.7}$As barrier layer grown on a GaAs (111)A substrate
by droplet epitaxy~\cite{Mano_APL10, Sallen_PRL11}. 
The formed QD has typically a truncated cone shape, and average dimension of the QD is
16 nm in radius and 1.4 nm in height.
Further details on the growth condition and the sample structure are
given elsewhere~\cite{Mano_APL10, Kuroda_PRB13}.
Since the (111) surface has $C_{3v}$ symmetry with identical in-plane covalent bonds,
one can expect mitigated QD's anisotropy~\cite{Singh_PRL09, Schliwa_PRB09, Juska_NPhot13},
which has well verified with atomic force microscopy analysis~\cite{Kuroda_PRB13}.
For optical characterization, sample was cooled to 9K
and a 640-nm pulsed semiconductor laser was used
to pump the Al$_{0.3}$Ga$_{0.7}$As barrier continuum.

\begin{figure}
\begin{center}
\includegraphics[width=18pc]{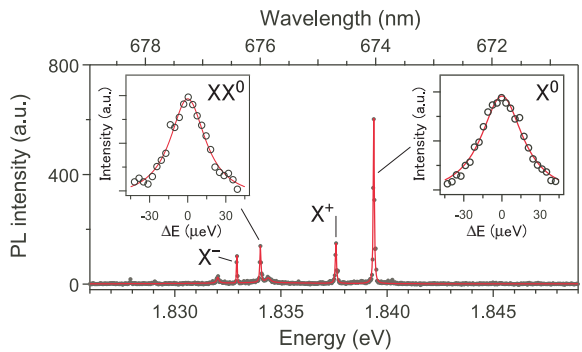}
\end{center}
\caption{PL spectrum of the isolated GaAs QD with a FSS
below the system resolution of 4 $\mu$eV measured at 9K.
Lorenzian fitting is also shown as a red curve.
Biphoton state is generated from $\left|{{\rm XX}^0} \right\rangle \to \left| {{\rm X}^0} \right\rangle \to \left| {\rm Vac} \right\rangle$ cascading process.
Expanded spectra are also displayed for ${\rm XX}^0$ and ${\rm X}^0$ lines.
}
\label{spectrum}
\end{figure}

Figure 1 shows the photoluminescence (PL) spectrum of an isolated GaAs dot.
In our samples, QDs have typically four emission lines,
i.e., negatively charged excitons (${\rm X}^-$),
neutral biexcitons (${\rm XX}^0$), positively charged excitons (${\rm X}^+$),
and neutral excitons (${\rm X}^0$) in order of increasing energy.
For performing photon correlation measurements, we have selected the as-grown QD
in which the FSS is below the system resolution of 4 $\mu$eV.
Biphoton correlation was investigated with a pair of cascadingly emitted ${\rm XX}^0$
and ${\rm X}^0$ photons by analyzing coincidence counts using time-to-digital converter (TDC).
In this work, we have measured coincidence with $6^2$ analyzer's polarization configurations,
i.e., $\left\{H, V, D, A, R, L \right\}$ polarizations for each line.
Figure 2 shows an example of accumulated coincidence counts for one analyzer configuration.

In order to analyze the biphoton state from the QD, density matrices were tomographically
reconstructed from the 36 datasets entailing maximally likelihood method~\cite{White_PRA01}.
Degree of mixedness and entanglement are evaluated in terms of
linear entropy (S$_L$)~\cite{Bose_PRA00}
and Tangle (T)~\cite{Wootters_PRL98}, respectively.
These measures can be calculated explicitly from the obtained density matrix $\rho$
as S$_L = \frac{4}{3}\left( {1 - {\rm{Tr}}{\rho^2}} \right)$,
and
T $= {\left[ {\max \left( {{\lambda _1} - {\lambda _2} - {\lambda _3} - {\lambda _4},0} \right)} \right]^2}$,
where $\lambda_i$ ($i$=1, 2, 3, 4)
is the square root of the eigenvalues in decreasing order of magnitude
of the spin-flipped density matrix operator
$R = {\rho }\left( {{\sigma ^y} \otimes {\sigma ^y}} \right)\rho^*\left( {{\sigma ^y} \otimes {\sigma ^y}} \right)$,
where $\sigma ^y$ is one of the Pauli's operators, and the asterisk indicates complex conjugation.
Biphoton states can be displayed in the S$_L-$T plane in Fig. 3(a),
in which all the physically allowed biphoton states will be mapped on the white region
below the dashed line.
In this plane, (S$_L$, T)=(0, 1) represents maximally entangled states,
while (S$_L$, T)=(1, 0) for totally mixed states.
To begin with, we analyzed the biphoton state without temporal gating.
In this case, outputs from the TDC were integrated over the range covering the whole peak,
(green stripe in the lower panel of Fig. 2)
and the resultant biphoton state is plotted as a green square in Fig. 3(a).
The state is on the Werner curve indicated by the solid line.
This is a clear manifestation that the emitted biphoton state from the QD is
maximally entangled mixed state for a given S$_L$ in the sense that
none of the unitary operations will restore the entanglement further~\cite{Ishizaka_PRA00}.
Another important consequence of this analysis is,
assuming the biphoton state being the Werner state,
one can readily see that the state is entangled for S$_L < 8/9$
and further violates local hidden variable model for S$_L < 1/2$~\cite{Kumano_OptExp11}.
In the present case, we have S$_L$=0.436 ($<$ 1/2) and T=0.382.
We can thus conclude the generated biphoton state from the present QD is nonlocal,
which is consistent with a direct demonstration of violating
Clauser, Horne, Shimony, and Holt (CHSH) version of the Bell's inequality~\cite{CHSH69}
without any fields nor filtering~\cite{Kuroda_PRB13}.

\begin{figure}
\begin{center}
\includegraphics[width=10pc]{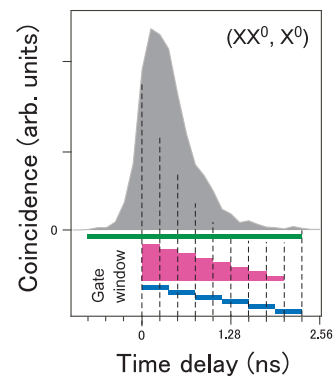}
\end{center}
\caption{(Upper panel) Typical lineshape of our histogram of the coincidence counts.
Displayed data was measured with (${\rm XX}^0$, ${\rm X}^0$) =(R, L) configuration.
(Lower panel) Integrating ranges for obtaining coincidence counts.
Green stripe covers whole peak with integration time of 3.072 ns,
and we refer to as {\it without} gating.
Two types of temporal gating, (i) widening gate (red) and (ii) shifting gate (blue)
are employed for evaluating the time evolution of the biphoton states
generated from a highly symmetric QD.
$\Delta t_g$ is 256 ps and 384 ps, respectively.
Time origin was determined to the point which gives the highest fidelity to
$\left| {{\Phi ^ + }} \right\rangle$.
Biphoton density matrices are tomographically reconstructed with 36 datasets
for each temporal gating.
}
\label{spectrum}
\end{figure}

The present highly symmetric QD emitter provides biphoton states
endowed with high degree of entanglement and nonlocality even without temporal gating.
This achievement sheds light on the dynamics underlying the intermediate exciton states,
such as coherent evolution of the state vector and the relaxation processes involved.
With a finite delay between a pair of ${\rm XX}^0$-${\rm X}^0$ photon generation
and the FSS (denoted by $S$),
possible biphoton pure state from the emitter is expressed as a bell state
$\left| {{\Phi ^ + }} \right\rangle = \frac{1}{{\sqrt 2 }}\left( {\left| {HH} \right\rangle + \left| {VV} \right\rangle} \right)$
with a relative phase of $\exp(iSt/\hbar)$ gained in the dwell time $t$
in the intermediate exciton state.
The probability of generating the state
from the QD emitter with an exciton lifetime of $\tau_r$, at the time between $t$ and $t+dt$,
is given by $\frac{1}{\tau_r} \exp(-t/\tau_r)$.
Therefore the biphoton state generated
in a time duration of $\left[t_g, t_g+\Delta t_g \right]$ is
\begin{eqnarray}
\hat{\rho} = \int_{t_g}^{t_g+\Delta t_g} \frac{1}{\tau_r} \exp(-t/\tau_r)\hat{\rho}_{\rm pure}(t)dt,\end{eqnarray}
where $\hat{\rho}_{\rm pure}(t)$ is the density matrix for the pure biphoton state
generated at the time $t$.
For constructing general biphoton density matrix,
we first consider effects of spin scattering and background (uncorrelated) light
by taking a convex combination of the $\hat{\rho}$ and totally mixed (uncorrelated) state of
$\frac{1}{4}I \otimes I$, where $I$ indicates identity operator of a single qubit.
The ratio of spin scattering (characteristic time $\tau_{ss}$) to radiative recombination
and the fraction of photon pairs stemming from the QD $k$ define the weight of $\hat{\rho}$
as $p \equiv k/(1+\tau_{r}/\tau_{ss})$.
%
In order to describe general biphoton mixed state from the QD including relaxation processes,
$1/\tau_r$ in the exponential function in Eq. 1 should be extended properly,
so that population with co-polarized components
($\left| {HH} \right\rangle $ and $\left| {VV} \right\rangle $)
decay with a rate of $1/\tau_{r}+1/\tau_{ss}$,
and decoherence takes place through further introduced parameter $\tau_{HV}$
in off-diagonal elements of the matrix
to characterize cross-dephasing~\cite{Hudson_PRL07, Add5}.
Thus we obtain the density matrix for the biphoton mixed state with the rectilinear bases of
$\left| {HH} \right\rangle,\left| {HV} \right\rangle,\left| {VH} \right\rangle,\left| {VV} \right\rangle$ as
\begin{equation}
\underline{\underline{\rho}} = \frac{1}{4}
\begin{pmatrix}
1+p & 0 & 0 & 2pI_c^{*}/I_0 \\
0 & 1-p& 0 & 0 \\
0 & 0& 1-p & 0 \\
2pI_c/I_0 & 0 & 0 & 1+p \\
\end{pmatrix}
,
\end{equation}
where
\begin{eqnarray}
I_0 = \int_{t_g}^{t_g+\Delta t_g} \frac{1}{\tau_r} e^{-t(1/\tau_r+1/\tau_{ss})}dt,
\end{eqnarray}
\begin{eqnarray}
I_c = \int_{t_g}^{t_g+\Delta t_g} \frac{1}{\tau_r} e^{-t(1/\tau_r+1/\tau_{ss}+1/\tau_{HV})}e^{iSt/\hbar}dt.
\end{eqnarray}
If $|I_c/I_0|=1$, hence $S=0$ and $p' \equiv k/(1+\tau_{r}/\tau_{ss}+\tau_{r}/\tau_{HV})=p$
(or equivalently $\tau_r/\tau_{HV}=0$),
the biphoton state $\underline{\underline{\rho}}$ reduces to the Werner state
and mapped onto an solid line in the S$_L-$T plane in Fig. 3(a) for $0 \le p' \le p \le 1$.
\begin{figure}
\begin{center}
\includegraphics[width=14pc]{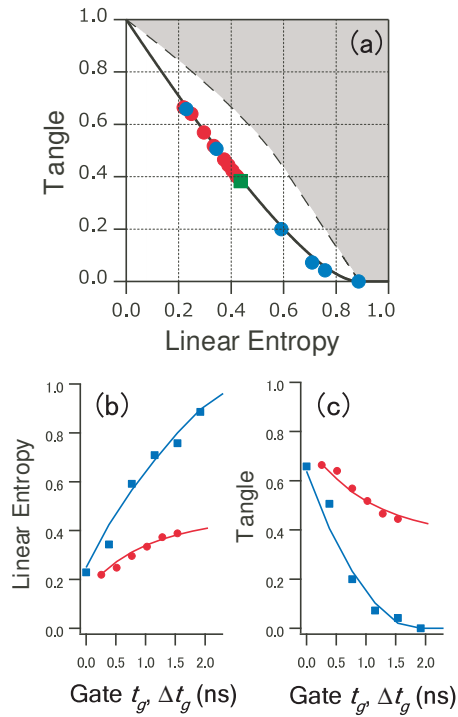}
\end{center}
\caption{Time evolution of the biphoton mixed state from a QD mapped on S$_L$-T plane.
Werner state is denoted by solid line.
(a) Experimentally obtained states via quantum tomography with 36 polarization basis
are shown by red (blue) circles employing widening (shifting) gate.
A green square exhibits the biphoton state without temporal gating.
The corresponding time evolution of the biphoton state in terms of
(b) linear entropy and (c) tangle is shown by solid lines.
Horizontal axis in (b) and (c) is $\Delta t_g$ $(t_g)$ for widening (shifting) gate.
}
\label{spectrum}
\end{figure}
In order to analyze the time evolution of the biphoton states from the QD,
two kind of temporal gating, i.e., (i) widening gate with constant time increment
and (ii) shifting gate with fixed width were employed in complementary manner.
Temporal gatings used for reconstructing the biphoton density matrices
are illustrated in the lower panel of Fig. 2.
For probing the coherent evolution, since the phase rotates
as the monitoring time extends, the widening gate can be more sensitive.
On the other hand, shifting gate is preferable to evaluate the relaxation dynamics.

In Fig. 3(a), experimentally reconstructed biphoton states
with the temporal gating are summarized as red (blue) circles
for the widening (shifting) gate.
By narrowing the gate width, the biphoton state moves toward maximally entangled
state (S$_L=0$) along the Werner curve.
With the narrowest gate of 256-ps width, we have S$_L$=0.219 and T=0.664.
For the shifting gate, biphoton state turned out
to evolve with time along the Werner curve to the totally mixed state (S$_L=1$)
with increasing the weight of mixed state component.
This finding indicates that the cross-dephasing is slow enough
in comparison to the radiative lifetime $(p'/p \simeq 1)$, 
and the phase rotation ${e}^{iSt/\hbar}$ contributes rather weakly to Eq. 4
$(S \simeq 0)$, which suggests that $S \ll \hbar/\Delta t_g$ = 1.7 $\mu$eV.
The time evolution of the biphoton state is also displayed in more explicit way
for both temporal gatings in terms of S$_L$ and T in Fig. 3(b) and (c), respectively.
In order to calculate the density matrix, the fraction of photon pairs exclusively
from the QD $k$ in $\underline{\underline{\rho}}$ is required.
For the analysis with temporal gating, since the $k$ depends on
the adopted gating condition,
we introduce alternative parameter $d$ to specify the ratio of uncorrelated (background)
count rate to the single rate for the QD emission at zero time delay~\cite{bg_parameter}.
By comparing the experimentally obtained biphoton states using independently measured parameters of $\tau_{r}$=560 ps and $\tau_{ss}$=2.8 ns for the identical dot,
we have found that
($S$, $\tau_{HV}$, $d$)
=(0.36 $\pm$ 0.06 $\mu$eV, 2.3 $\pm$ 0.5 ns, 0.008 $\pm$ 0.004)
gives the best agreement to the experimental observation in Fig. 3(a)-(c).
Basically these parameters were obtained
to reproduce the upper and lower bounds for the biphoton state
with shifting and widening gates in Fig. 3(a) and overall behavior
in Fig. 3(b) and (c).
Since the density matrix has full information on the biphoton state,
we can deduce the fundamental parameters to characterize the exciton dynamics
by analyzing the matrix as a function of the delay time
between ${\rm XX}^0$ and ${\rm X}^0$ photo-generation.

\begin{figure}
\begin{center}
\includegraphics[width=9pc]{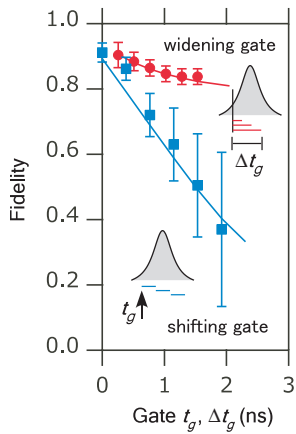}
\end{center}
\caption{Fidelity $f$ to the $\left| {{\Phi ^ + }} \right\rangle  = \frac{1}{{\sqrt 2 }}\left( {\left| {HH} \right\rangle + \left| {VV} \right\rangle} \right)$
for the experimentally obtained biphoton states with widening gate (red circles)
and shifting gate (blue squares). Error bars represent one standard deviation.
Analytically calculated fidelity employing the identical parameters with Fig. 3
is also displayed as solid lines for each gating conditions.
$f \simeq (1+p)/4 + p^2/2p'$ for zero gate limit.
}
\label{spectrum}
\end{figure}

In Fig. 4, entanglement fidelity $f$ of the experimentally obtained biphoton states
to the maximally entangled state
$\left| {{\Phi ^ + }} \right\rangle$ is examined.
The fidelity is calculated with $f=(1+C_{H/V}+C_{D/A}+C_{R/L})/4$
for the two temporal gating conditions, where $C_{H/V}$, $C_{D/A}$, and $C_{R/L}$ is a correlation
function in rectilinear, diagonal, and circular basis, respectively.
%
%
Overall behavior agrees well with
$\left\langle {\Phi ^ + } \right|{\underline{\underline{\rho}}}\left| {\Phi ^ + }
\right\rangle = (1+p+2p'{\rm Re}(I_c/I_0))/4$
using the identical parameters in Fig. 3(b),(c) (solid lines),
which indicates that the analytical method presented in this letter
is quite useful to grasp the comprehensive understandings of the biphoton state from the QD.
It is noteworthy that even for a zero gate width limit,
the fidelity is below unity.
This is due to residual non-zero mixedness given by S$_L \simeq (-2p^4/p'^2-p^2+3)/3$.
In order to realize ideal biphoton pure sources with S$_L \simeq 0$,
we need $\tau_{r}/\tau_{ss} \simeq 0$, $\tau_{r}/\tau_{HV} \simeq 0$, and $k \simeq 1$.
Shortening of $\tau_{r}$ by introducing Purcel effect~\cite{Purcell46}
with background-free QDs and stabilizing spin states
in the intermediate $\left| {{\rm X}^0} \right\rangle$ state will be crucial
toward this direction.

In conclusion
we have demonstrated "as-grown" nonlocal biphoton generation from a QD
without applying fields or gatings.
The biphoton state generated from a QD can be well described as
Werner state, which discriminates the state against generally allowed biphoton mixed states.
Since the Werner state is the maximally entangled mixed biphoton state,
its generation from the solid-state quantum dot in deterministic manner
will be quite beneficial to the versatile fields of quantum information applications.
We have also found that the cross-dephasing which loses coherence between two pathways
in cascading process $\left|{{\rm XX}^0} \right\rangle \to \left| {{\rm X}^0} \right\rangle \to \left| {\rm Vac} \right\rangle$
occurs less frequently than the exciton radiative recombination.

This work was supported in part by JSPS KAKENHI Grant Number 24310084
and the Murata Science Foundation.

\bibliography{kuma_PRL}

\begin{thebibliography}{43}
\bibitem [{add()}]{add3}%
  \BibitemOpen
  \href@noop {} {}\bibinfo {note} {Originally, convex combination of $\left|
  {{\Psi ^ - }} \right\rangle = \frac{1}{{\sqrt 2 }}\left({\left| {HV}
  \right\rangle - \left| {VH} \right\rangle} \right)$ and $I \otimes I$ is
  called Werner state, which is invariant under arbitrary unitary operation for
  both qubits $U \otimes U$. However, four Bell states $\left| {{\Psi ^ {\pm}
  }} \right\rangle$ and $\left| {{\Phi ^ {\pm} }} \right\rangle$ are
  interchangable with local operations and the resultant states exhibit the
  same entanglement properties. Thus $\left| {{\Phi ^ + }} \right\rangle$ mixed
  with $I \otimes I$ is also referred to as the Werner state.}\BibitemShut
  {Stop}%
\bibitem [{Add()}]{Add5}%
  \BibitemOpen
  \href@noop {} {}\bibinfo {note} {Cross dephasing in a QD is equivalent to
  dichroic scatterers for PDC sources as discussed in ref. 18}\BibitemShut
  {NoStop}%
\bibitem [{bg_()}]{bg_parameter}%
  \BibitemOpen
  \href@noop {} {}\bibinfo {note} {We assume $d$ is constant as a function of
  delay time, and is related to $k$ as $k = \left[ \frac{S(t_1, t_2)}{S(t_1,
  t_2)+B(t_1, t_2)} \right]^2$, where $S(t_1, t_2) = \int_{t_1}^{t_2}
  \exp(-t/\tau_r)dt$, and $B(t_1, t_2) = d(t_2-t_1)$.}\BibitemShut {Stop}%
\end{thebibliography}

\end{document}